\documentstyle[preprint,eqsecnum,cite,epsf,aps,amssymb]{revtex}
\newcommand{\be}{\begin{equation}}
\newcommand{\ee}{\end{equation}}
\newcommand{\bea}{\begin{eqnarray}}
\newcommand{\eea}{\end{eqnarray}}
\newcommand{\ba}{\begin{array}}
\newcommand{\ea}{\end{array}}
\newcommand{\nn}{\nonumber \\}
\newcommand{\half}{\frac{1}{2}}

\begin{document}
\tightenlines

\title{Noncommutative Supersymmetric Field Theories \footnote{Updated
    version of a talk presented at the XXI Brazilian National Meeting
    on Particles and Fields, S\~ao Loren\c co, MG, Oct. 23 - 26, 2000}}

\author{Victor O. Rivelles}

\address{Instituto de F\'\i sica, Universidade de S\~ao Paulo\\
 Caixa Postal 66318, 05315-970, S\~ao Paulo - SP, Brazil\\
E-mail: rivelles@fma.if.usp.br}

\maketitle

\begin{abstract}
We discuss some properties of noncommutative supersymmetric field
theories which do not involve gauge fields. We concentrate on the
renormalizability issue of these theories. 
\end{abstract}

\section{Introduction}
Although string theory is quite well understood in the perturbative
regime its formulation in a background independent way is almost
unknown. There are many reasons for that. String theory has too many
degrees of freedom. It is quite difficult to handle all of them
together. It also includes the  
gravitational field which may have quantum fluctuations. And there are
many sources for nonlocality which is also troublesome in any
theory. One way out of these difficulties is to
consider limits of string theory which have some of the troubles 
raised above but not all of them. This may allow us to understand
better some aspects of string theory without the complications of the
full theory. 

One such a limit is the zero slope limit of the D3-brane in the presence
of a constant NS-NS field \cite{Seiberg-Witten}. The low energy
effective theory is a quantum field theory deformed in terms of the Moyal
product over space-time. In noncommutative field theories the usual
product of fields is replaced by the Moyal product of fields giving
rise to nonlocal field theories  \cite{Filk}. Usually nonlocal field
theories turn out to be not well 
defined but the nonlocality induced by the Moyal product is still
tractable. It was found that the main characteristic of
noncommutative field theories is the mixing of ultraviolet (UV) and
infrared (IR) divergences due to its nonlocal structure
\cite{Minwalla}. As a 
consequence it is not clear that the properties of the usual
commutative field theories are kept, without modifications, in their  
noncommutative counterparts. This gave rise to an intensive research of
noncommutative field theories in Euclidean or Minkowski
space-time.

One of the manifestations of the UV/IR mixing in the $\lambda
\phi^4$ theory is as an infrared quadratic singularity in the
propagator at one loop \cite{Minwalla}. Although renormalizable up to
two loops \cite{Arefeva} it 
becomes non-renormalizable at higher loop orders. Models involving a
complex scalar field may be non-renormalizable even at one loop
\cite{Arefeva-complex}. So, 
noncommutativity seems to destroy the main characteristic of commutative
field theories, i.e., their renormalizability. 

In what follows we will discuss the inclusion of supersymmetry in such
models and how it restores the renormalizability. We will concentrate
on the Wess-Zumino model in $3+1$ dimensions \cite{Wess-Zumino} and
the supersymmetric non-linear sigma model in $2+1$ dimensions
\cite{Sigma-Model}. In this last case we will 
see that the noncommutativity also destroys the mechanism for dynamical mass
generation of the fermionic sector, and we will show how supersymmetry
helps to fix it. 

\section{Noncommutative Spaces}

In quantum mechanics we have the usual commutation relations 
\bea
\label{1.1}
\left[ \hat{q}^i, \hat{p}^j \right] &=& i \hbar g^{ij}, \\
\left[ \hat{q}^i, \hat{q}^j \right] &=& \left[ \hat{p}^i, \hat{p}^j
  \right] = 0.
\eea
It is natural to consider noncommutative coordinates with commutation
relations
\be 
\label{2.1}
\left[ \hat{q}^i, \hat{q}^j \right] = i \theta^{ij},
\ee
where $\theta_{ij}$  is a constant of dimension $L^2$ which defines a
noncommutativity scale. This breaks
rotational (or Lorentz) symmetry but in the limit $\theta \rightarrow
0$ the symmetry is recovered. This is an example of a noncommutative
space. It can be extended to space-time but we will consider
noncommutativity only in the spatial coordinates since otherwise there are
problems with unitarity \cite{Seiberg-Susskind}. 

We can understand heuristically how the UV and IR physics gets 
mixed. From Eq.(\ref{1.1}) it follows that $\Delta \hat{q}^i 
\Delta \hat{p}^j \sim i g^{ij}$. In a similar way, from Eq.(\ref{2.1})
it follows that $\Delta \hat{q}^i \Delta \hat{q}^j \sim i
\theta^{ij}$ so we expect that $\Delta \hat{q} \sim \theta \Delta
\hat{p}$. This means that 
high energy modes have drastic effects at large distances (or small
energy processes). As we shall see, in quantum field theory this
mixing manifests itself already at one loop level in the propagator of
the fields. 

Fields defined on such spaces are operator valued objects. 
It turns out to be more convenient to use fields which are not
operator valued objects but just functions. This can be achieved
through the use of the Weyl-Moyal correspondence \cite{Filk}
\be
\hat{\Phi}(\hat{q}) \rightarrow \phi(x).
\ee
We
associate to the operator valued field $\hat{\Phi}(\hat{q})$ a
classical function $\phi(x)$ through its Fourier transform
$\tilde{\phi}(p)$ as 
\be
\hat{\Phi}(\hat{q}) = \int dp \,\, e^{i p \hat{q} } \tilde{\phi}(p).
\ee
The operator valued field $\hat{\Phi}$ satisfies 
\be
\hat{\Phi}_1(\hat{q}) \hat{\Phi}_2(\hat{q}) = \int dp_1 \,\, dp_2 \,\,
e^{i(p_1 + p_2) \hat{q} - \frac{1}{2}  p_1^\mu p_2^\nu
  \theta_{\mu\nu}} \phi_1(p_1) \phi_2(p_2),
\ee
hence
\be
\hat{\Phi}_1(\hat{q}) \hat{\Phi}_2(\hat{q}) \leftrightarrow \left( \phi_1
  \star \phi_2 \right) (x),
\ee
where
\be
\label{4}
\left( \phi_1 \star \phi_2 \right) (x) \equiv \left[ e^{i\frac{1}{2}
    \theta^{\mu\nu} \frac{\partial}{\partial x^\mu}
    \frac{\partial}{\partial y^\nu} } \phi_1(x) \phi_2(y)
\right]_{y=x},
\ee
is the Moyal (or star) product. Then we can work on a commutative
space in which the  usual product of field is replaced by the Moyal
product. Notice that the derivatives in the definition Eq.(\ref{4})
    makes the Moyal product non-local. Also, the Moyal commutator of
    the commutative coordinates $x^\mu$ gives
\be 
\left[ x^\mu, x^\nu \right]_{MB} = x^\mu \star x^\nu - x^\nu \star
    x^\mu = i \theta^{\mu\nu}.
\ee

It can be easily verified the following properties of the Moyal
product:
\be
a) \qquad e^{ikx} \star e^{iqy} = e^{i(k+q)x} e^{-i k \wedge q},
\ee
where $k \wedge q = \half k^\mu \theta_{\mu\nu} q^\nu$.
\be
b) \qquad (f \star g)(x) = \int dk \,\, dq \,\, \tilde{f}(k)
\tilde{g}(q) e^{-ik   \wedge q} e^{i (k+q)x},
\ee
where $\tilde{f}$ and $\tilde{g}$ are the Fourier components of $f$
and $g$, respectively. 
\be 
c) \qquad \left[ (f \star g) \star h \right] (x) = \left[ f \star ( g
  \star h ) \right] (x).
\ee
\be
d) \qquad \int dx \,\, (f \star g)(x) =  \int dx \,\, (g \star f)(x) =
\int dx \,\, f(x) g(x).
\ee
\be 
e) \qquad \int dx \,\, (f_1 \star f_2 \star \dots f_n)(x) =  \int dx
\,\, (f_n \star f_1 \star \dots f_{n-1})(x).
\ee
\be
f) \qquad (f \star g)^* = g^* \star f^*.
\ee

\section{Noncommutative Scalar Field Theory}

Let us consider the massive scalar field in $D=3+1$ dimensions
\cite{Minwalla}, whose action is 
\be
S = \int d^4 x \,\, \left( \half \partial_\mu \phi \star \partial^\mu \phi
- \frac{m^2}{2} \phi \star \phi - \frac{g^2}{4!}
\phi\star\phi\star\phi\star\phi \right).
\ee
Using property d) it is seen that the propagator is not affected by
the Moyal product. This is a generic property of noncommutative field
theories. The vertex, however, must be symmetrized . In
momentum space we have
\bea
&& -\frac{g^2}{2} \int d^4x \,\, \phi\star\phi\star\phi\star\phi =
-\frac{g^2}{6} \int dk_1 dk_2 dk_3 dk_4 \,\, \delta 
(k_1+k_2+k_3+k_4) \times \nn
&& [ \cos(\half k_1 \wedge k_2) \cos (\half k_3 \wedge k_4) + 
\cos(\half k_1 \wedge k_3) \cos (\half k_2 \wedge k_4) + \nn
&& \cos(\half k_1 \wedge k_4) \cos (\half k_2 \wedge k_3) ] \,\,
\phi(k_1) \, \phi(k_2) \,  \phi(k_3) \, \phi(k_4). 
\eea

Then, the one loop correction for the two-point function is
\be 
\frac{g^2}{3 (2 \pi)^4} \int d^4k \,\, \left( 1 + \half \cos(k \wedge
  p) \right) \frac{1}{k^2 + m^2}.
\ee
The first term is the usual one loop mass correction of the
commutative theory (up to a factor $1/2$) which is quadratically
divergent. The second term is not divergent due to the oscillatory
nature of $\cos(k \wedge 
p)$. This shows that the nonlocality introduced by the Moyal product
is not bad and leaves us with the same divergence structure of the
commutative theory. To take into account the effect
of the second term we regularize the integral using the Schwinger
parametrization 
\be
\frac{1}{k^2 +m^2} = \int_0^\infty d\alpha \,\, e^{- \alpha (k^2 +
  m^2)} e^{-\frac{1}{\Lambda^2 \alpha}},
\ee
where a cutoff $\Lambda$ was introduced. We find
\be \Gamma^{(2)} = \frac{g^2}{48 \pi^2} [ ( \Lambda^2 - m^2
\ln(\frac{\Lambda^2}{m^2}) + \dots ) + \half ( \Lambda^2_{eff} - m^2
\ln(\frac{\Lambda^2_{eff}}{m^2}) + \dots ) ],
\ee
where
\be
\Lambda^2_{eff} = \frac{1}{\frac{1}{\Lambda^2} + \tilde{p}^2}, \qquad
\tilde{p}^\mu = \theta^{\mu\nu} p_\nu.
\ee
Note that when the cutoff is removed, $\Lambda \rightarrow \infty$,
the noncommutative contribution remains finite providing a natural
regularization. Also $\Lambda^2_{eff} = \frac{1}{\tilde{p}^2}$ which
diverges either when  $\theta \rightarrow 0$ or when $\tilde{p}
\rightarrow 0$. 

The one loop effective action is then
\be
\int d^4p \,\, \half ( p^2 + M^2 + \frac{g^2}{96 \pi^2
(\tilde{p}^2 + 1/\Lambda^2)} - \frac{g^2 M^2}{96 \pi^2}
\ln\left(\frac{1}{M^2(\tilde{p}^2 + 1/\Lambda^2)} \right) + \dots )
  \phi(p) \phi(-p),
\ee
where $M$ is the renormalized mass. Let us take the limits $\Lambda
\rightarrow \infty$ and $\tilde{p} \rightarrow 0$. If we take first
$\tilde{p} \rightarrow 0$ then $\tilde{p}^2 << \frac{1}{\Lambda^2}$
and $\Lambda_{eff}=\Lambda$ showing that we 
recover the effective commutative theory
\be
\int d^4p \,\, \half \left( p^2 + M^{\prime 2} \right) \phi(p)
\phi(-p). 
\ee 
If, however, we take $\Lambda \rightarrow \infty$ then $\tilde{p}^2 >>
\frac{1}{\Lambda^2}$ and $\Lambda^2_{eff} = \frac{1}{\tilde{p}^2}$
and we get  
\be 
\int d^4p \half \left( p^2 + M^2 + \frac{g^2}{96 \pi^2 \tilde{p}^2} -
  \frac{g^2 M^2}{96 \pi^2} \ln\left(\frac{1}{M^2 \tilde{p}^2}\right) +
  \dots \right) \phi(p) \phi(-p),
\ee
which is singular when $\tilde{p} \rightarrow 0$. 
This shows that the limit $\Lambda \rightarrow \infty$ does not
commute with the low momentum limit $\tilde{p} \rightarrow 0$  so that
there is a mixing of UV and IR limits. 

The theory is renormalizable at one loop order if we do not
take $\tilde{p} \rightarrow 0$. What about higher loop orders? Suppose
we have insertions of one loop mass corrections. Eventually we will have
to integrate over small values of $\tilde{p}$ which diverges when
$\Lambda \rightarrow \infty$. Then we find an IR
divergence in a massive theory. This combination of UV and IR
divergences makes the theory non-renormalizable. 

There are also examples of non-renormalizable theories already at
one loop order \cite{Arefeva-complex}. For a complex scalar field with
interaction  $\phi^*\star \phi^* \star \phi \star \phi$ it is found
that the theory is one-loop non-renormalizable while $\phi^*\star \phi
\star \phi^*\star \phi$ gives a  one loop renormalizable  model. 

Then the question is whether it would be possible to find a theory
which is renormalizable to all loop orders. Since the UV/IR mixing
appears at the level of quadratic divergences a candidate theory would
be a supersymmetric theory because it does not have such divergences
\cite{Chepelev,Ferrara}. As we shall see this indeed happens. 

\section{NONCOMMUTATIVE WESS-ZUMINO MODEL}

The noncommutative Wess-Zumino model in $3+1$ dimensions
\cite{Wess-Zumino} has the action 

\bea
{\cal L}_0 &=&  \frac12 \partial^\mu A \partial_\mu A + \frac12
\partial^\mu B \partial_\mu B
+ \frac12 \overline \psi i\not \! \partial \psi, \\
{\cal L}_m &=& \frac12 F^2+\frac12 G^2
+ m F A + m G B - \half m \overline\psi \psi, \\ 
{\cal L}_g &=&  g (F\star A \star
A- F\star B \star B + G\star A \star B + G \star B \star A - \nn
&& \overline \psi
\star \psi\star A - \overline \psi\star  i\gamma_5 \psi \star B),
\eea
where $A$ and $B$ are bosonic fields, $F$ and $G$ are auxiliary
fields and $\psi$ is a Majorana spinor. The action is invariant under
the usual supersymmetry transformations. They are not modified by the 
Moyal product since they are linear in the fields. The elimination of
the auxiliary fields through their equations of motion produces quartic
interactions. In terms of the complex field $\phi=A+iB$ we get $\phi^*
\star \phi^* \star \phi \star \phi$ which is non-renormalizable in the
noncommutative case. This casts doubts about the renormalizability of the
model but as we shall see supersymmetry saves the day. 

As usual, the propagators are not modified by noncommutativity due to
the property d). They are given by
\bea
\Delta_{AA}(p) &=& \Delta(p)\equiv\frac{i}{p^2-m^2+i\epsilon},\\
\Delta_{FF}(p) &=& p^2 \Delta(p),\\
\Delta_{AF}(p) &=&\Delta_{FA}(p) = -m \Delta(p), \\
S(p) &=& \frac{i}{\not \! p -m}.
\eea
Taking into account the symmetries the vertices are 
\bea
F A^2 \quad {\mbox {vextex:}}&& \quad ig \cos(p_1\wedge p_2), \\
F B^2  \quad {\mbox {vextex:}}&& \quad -ig \cos(p_1\wedge p_2),\\
G A B \quad  {\mbox {vertex:}} && \quad 2 ig \cos (p_1\wedge p_2),\\
\overline \psi \psi A\quad {\mbox {vertex:}} &&\quad -ig 
\cos (p_1\wedge p_2),\\ 
\overline \psi \psi B\quad {\mbox {vertex:}} &&\quad -ig\gamma_5 
\cos (p_1\wedge p_2).
\eea
The degree of superficial divergence for a generic 1PI graph
$\gamma$ is then 
\be
d(\gamma)= 4 -  I_{AF} -I_{BF}-N_A-N_B-2 N_F-2N_G - \frac32 N_\psi, 
\ee
where $N_{\cal O}$ denotes the number of external lines
associated to the  field ${\cal O}$ and $ I_{AF}$ and $I_{BF}$
are the numbers of internal lines associated to the mixed
propagators $AF$ and $BF$, respectively. In all cases we will
regularize the divergent Feynman integrals by assuming that a 
supersymmetric regularization scheme does exist.

The one loop analysis can be done in a straightforward way. 
As in the commutative case all tadpoles contributions add up to
zero. We have verified this explicitly. The self-energy of $A$ can
be computed and the divergent part is contained in the integral
\be
16 g^2\int\frac{d^4k}{(2\pi)^4}( 1 + \half \cos(k \wedge p) ) \frac{(p \cdot
  k)^2}{(k^2- m^2)^3}.
\ee
The first term is logarithmically divergent. It differs by a factor 2
from the commutative case. As usual, this 
divergence is eliminated by a wave function renormalization. 
The second term is UV convergent and for small $p$ it behaves as 
$p^2 \ln (p^2/m^2)$ and actually vanishes for $p=0$. Then there is no
IR pole. The same analysis can be carried out for the others
fields. For $F$ we find that the divergent part is
\be
4 g^2 \int \frac{d^4k}{(2\pi)^4 } (1 + \half\cos(k \wedge p) )
\frac{1}{(k^2-m^2)^2}.
\ee
The first term is logarithmically divergent and can also be eliminated
by a wave 
function renormalization. The second term diverges as $\ln(p^2/m^2)$
as $p$ goes to zero. However its multiple insertions is harmless. 
For the fermion field the divergent part is similar to the former
results and needs also a wave
function renormalization. The term containing $\cos(k\wedge p)$ 
behaves as $\not \! p \ln(p^2/m^2)$ and vanishes as $p$ goes to
zero. Therefore, there is no UV/IR mixing in the self-energy as
expected. 

To show that the model is renormalizable we must also look into the
interactions vertices. The $A^3$ vertex has no divergent parts as in
the commutative case. The same happens for the other three point
functions. For the four point vertices no divergence is found as in
the commutative case. Hence, the noncommutative Wess-Zumino model is
renormalizable at one loop with a wave-function renormalization and no
UV/IR mixing.  

To go to higher loop orders we proceed as in the commutative case
\cite{Iliopoulos}. We derived the supersymmetry Ward identities 
for the n-point vertex function. Then we showed that there is a
renormalization prescription which is consistent with the Ward
identities. They are the same as in the commutative case. And finally
we fixed the primitively divergent vertex functions. Then we found
that there is only a common wave function renormalization as in the
commutative case. In general we expect 
\be 
\varphi_R = Z^{-1/2} \varphi, \qquad m_R = Z m + \delta m, \qquad
g_R = Z^{3/2} Z^\prime g.
\ee
At one loop we found { $\delta m = 0$ and $Z^\prime = 1$}. We showed
that this also holds to all orders and no mass renormalization is
needed. 

Being the only consistent noncommutative quantum field theory in $3+1$
dimensions known so far it is natural to study it in more detail. As a
first step in this direction we considered the nonrelativistic limit
of the noncommutative Wess-Zumino model \cite{low_energy}. We found
the low energy 
effective potential mediating the fermion-fermion and boson-boson
elastic scattering in the nonrelativistic regime. Since
noncommutativity breaks Lorentz invariance we formulated the theory in
the center of mass frame of reference where the dynamics simplifies
considerably. For the fermions we found that the potential is
significatively changed by the noncommutativity while no modification
was found for the bosonic sector. The modifications found give rise to
an anisotropic differential cross section. 

\section{Noncommutative Gross-Neveu and  Nonlinear Sigma Models}

Another model where nonrenormalizability is spoiled by the
noncommutativity is the $O(N)$ Gross-Neveu model. This model is
perturbatively renormalizable in $1+1$ dimensions and $1/N$
renormalizable in $1+1$ and $2+1$ dimensions. In both cases it
presents dynamical mass generation. It is described by the Lagrangian 
\begin{equation}
{\cal L}=\frac{i}2\overline \psi_i \not \!\partial \psi_i +\frac
{g}{4N}(\overline \psi_i \psi_i)(\overline \psi_j \psi_j),\label{1}
\end{equation}
where $\psi_i, i=1,\ldots N$, are two-component Majorana
spinors. Since it is renormalizable in the $1/N$ expansion in $1+1$
and $2+1$ dimensions we will consider both cases. As usual, we
introduce an auxiliary field $\sigma$ and the Lagrangian turns into 
\begin{equation}
{\cal L} =\frac{i}2\overline \psi_i \not \!\partial \psi_i -\frac{\sigma}2 
(\overline \psi_i \psi_i)- \frac{N}{4g}\sigma^2.\label{2}
\end{equation}
Replacing $\sigma$ by $\sigma+M$ where $M$ is the VEV of the
original $\sigma$ we get the gap equation (in Euclidean space) 
\begin{equation}
 \frac{M}{2g}-\int \frac{d^Dk}{(2\pi)^D}\frac{M}{k^{2}_{E}+M^2}=0.\label{5}
\end{equation}
To eliminate the UV divergence we need to renormalize the coupling
constant by 
\begin{equation}
\frac{1}{g}= \frac{1}{g_R} + 2 \int \frac{d^Dk}{(2\pi)^D}\frac{1}{k^{2}_{E}
+\mu^2}. \label{6}
\end{equation}
In $2+1$ dimensions we find 
\begin{equation}
\frac{1}{g_R}= \frac{\mu-|M|}{2\pi}, \label{7}
\end{equation}
and therefore only for $-\frac{1}{g_R}+\frac{\mu}{2\pi}>0$ it is possible to
have $M\not =0$, otherwise $M$ is necessarily zero. 
No such a restriction exists in $1+1$ dimensions. In any case, 
we will focus only in the massive phase.
The propagator for $\sigma$ is proportional to the inverse of the
following expression
\begin{equation}
 -\frac {iN}{2g}-  iN \int \frac{d^Dk}{(2\pi)^D} \frac{k\cdot (k+p) + 
M^2}{(k^2-M^2)[(k+p)^2-M^2]}, \label{8}
\end{equation}
which is divergent. Taking into account the gap equation the above
expression reduces to 
\begin{equation}
\frac{(p^2-4M^2)N}2\int\frac{d^Dk}{(2\pi)^D} \frac{1}{(k^2-M^2)[(k+p)^2-M^2]},
\label{10}
\end{equation}
which is finite. Then there is a fine tuning which is responsible for
the elimination of the divergence and which might be absent in the
noncommutative case due to the UV/IR mixing. 

The noncommutative model is defined by
\begin{equation}
S_{GN}=\int d^Dx \left [\frac{i}2\overline \psi \not \!\partial \psi - \frac{M}2\overline \psi
\psi-\frac12\sigma \star(\overline \psi\star
\psi)- \frac{N}{4g}\sigma^2- \frac{N}{2g}M\sigma\right ].\label{11}
\end{equation} 
Elimination of the auxiliary field results in a four-fermion
interaction of the type $\overline \psi_i\star\psi_i\star\overline
\psi_j\star\psi_j$. However a more general four-fermion interaction
may involve a term like $\overline \psi_i\star\overline
\psi_j\star\psi_i\star\psi_j$. This last combination does not have a
simple $1/N$ expansion and we will not consider it. The Moyal product
does not affect the propagators and the trilinear vertex acquires a
correction of $\cos(p_1\wedge p_2)$ with regard to the commutative
case. Hence the gap equation is not modified, while the propagator for
the $\sigma$ is now proportional to the inverse of
\begin{equation}
 -\frac {iN}{2g}- N \int \frac{d^Dk}{(2\pi)^D} \cos^2(k\wedge p)
\frac{k\cdot (k+p) + 
M^2}{(k^2-M^2)[(k+p)^2-M^2]}. \label{12}
\end{equation}
Now the divergent part is no longer canceled and this turns the model
into a nonrenormalizable one. 

On the other side, the nonlinear sigma model also presents troubles in
its noncommutative version. The noncommutative model is described by 
\begin{equation}\label{121}
{\cal L} =-\frac12\varphi_i (\partial^2 + M^2)\varphi_i + \frac12
\lambda\star \varphi_i \star \varphi_i - \frac{N}{2g}\lambda,
\end{equation}
where $\varphi_i$, $i=1,\ldots , N$, are real scalar fields, $\lambda$
is the auxiliary field and $M$ is the generated mass. The leading
correction to the $\varphi$ self-energy is
\begin{equation}\label{122}
-i\int \frac{d^2 k}{(2\pi)^2} \frac{\cos^2 (k\wedge p)}{(k+p)^2-M^2}\Delta_\lambda(k),
\end{equation}
where $\Delta_\lambda$ is the propagator for $\lambda$. As for the
case of the scalar field this can be decomposed as a sum of a
quadratically divergent part and a UV finite part. Again there is the
UV/IR mixing destroying the $1/N$ expansion. 

\section{Noncommutative Supersymmetric Nonlinear Sigma Model}

The Lagrangian for the commutative supersymmetric sigma model is given
by
\begin{equation}
{\cal L} =\frac12 \partial^\mu \varphi_i \partial_\mu \varphi_i +
\frac{i}{2} \overline \psi_i \not \! \partial \psi_i + \frac12 F_i F_i
+ \sigma \varphi_i F_i + \frac12\lambda 
\varphi_i \varphi_i  - \frac12\sigma \overline \psi_i \psi_i -
\overline \xi \psi_i \varphi_i - \frac{N}{2g}\lambda,\label{13} 
\end{equation}
where $F_i$, $i=1,\ldots, N$, are auxiliary fields. Furthermore,
$\sigma,\lambda$ and $\xi$ are the Lagrange multipliers which
implement 
the supersymmetric constraints. After the change of variables 
$\lambda\rightarrow \lambda + 2 M \sigma$, $F\rightarrow F-M\varphi$
where $M=<\sigma>$, and the shifts $\sigma\rightarrow \sigma +M$ and
$\lambda\rightarrow \lambda + \lambda_0$, where $\lambda_0=<\lambda>$,
we arrive at a more symmetric form for the Lagrangian 
\begin{eqnarray}
{\cal L} &=& - \frac12 \varphi_i ( \partial^2+M^2) \varphi_i + \frac{1}{2} 
\overline \psi_i (i\not \! \partial - M)\psi_i + \frac12 F_i^2+
 M^2 \varphi_i^2+ \frac12\lambda_0 \varphi_i^2   \nonumber\\ 
&\phantom a& + \frac12\lambda \varphi_i^2 +\sigma  \varphi_i F_i
 - \frac12\sigma \overline \psi_i \psi_i  - \overline \xi \psi_i
 \varphi_i- \frac{N}{2g}\lambda -\frac{N}{g}M\sigma.\label{15} 
\end{eqnarray}
Now supersymmetry requires $\lambda_0=-2M^2$ and the gap equation is 
\begin{equation}
\int \frac{d^D k}{(2\pi)^D} \frac{i}{k^2-M^2}= \frac{1}{g},\label{16}
\end{equation}
so a coupling constant renormalization is required. We now must
examine whether the propagator for $\sigma$ depends on the this
renormalization.  We find that the two point function for $\sigma$ is
proportional to the inverse of
\begin{equation}
 \frac{(p^2-4M^2)N}2\int\frac{d^Dk}{(2\pi)^D}
 \frac{1}{(k^2-M^2)[(k+p)^2-M^2]}\,\,\label{18},
\end{equation}
which is identical to the Gross-Neveu case. Notice that the gap equation
was not used. The finiteness of the above expression is a consequence
of supersymmetry. 

The noncommutative version of the supersymmetric nonlinear sigma model
is given by 
\begin{eqnarray}
{\cal L} &=&  
- \frac12\varphi_i (\partial^2+M^2) \varphi_i + \frac{1}{2} \overline
\psi_i (i\not \! \partial -M)\psi_i + \frac12 F_i^2 +
\frac{\lambda}{2}\star\varphi_i \star\varphi_i  \nonumber \\
& \phantom a & - \frac12 F_i \star (\sigma\star\varphi_i +\varphi_i
\star\sigma)-\frac12 \sigma\star\overline\psi_i \star \psi_i  
-\frac12 (\bar\xi\star\psi_i \star\varphi_i
+\bar\xi\star\varphi_i\star\psi_i ) \nonumber\\
&\phantom a &-\frac{N}{2g}\lambda - \frac{N M\sigma}{g}.
\label{19}
\end{eqnarray}
Notice that supersymmetry dictates the form of the trilinear
vertices. Also, the supersymmetry transformations are not modified by
noncommutativity since they are linear and no Moyal products are
required. 

The propagators are the same as in the commutative case. The vertices
have cosine factors due to the Moyal product 
\begin{mathletters}
\begin{eqnarray}\label{20}
\lambda \varphi^2 \quad {\mbox {vertex:}}&& \quad \frac{i}2 
\cos(p_1\wedge p_2), \\
\sigma \varphi F \quad  {\mbox {vertex:}} && \quad - i \cos (p_1\wedge p_2),\\
\overline \psi \psi \sigma \quad {\mbox {vertex:}} &&\quad -\frac{i}2 
\cos (p_1\wedge p_2),\\ 
\overline \xi \psi \varphi \quad {\mbox {vertex:}} &&\quad - i 
\cos (p_1\wedge p_2).
\end{eqnarray}
\end{mathletters}
We again consider the propagators for the Lagrange multiplier
fields. Now the $\sigma$ propagator is modified
by the cosine factors and is proportional to the inverse of 
\begin{equation}
\frac{(p^2-4M^2)N}2\int\frac{d^Dk}{(2\pi)^D}
 \frac{\cos^2(k\wedge p)}{(k^2-M^2)[(k+p)^2-M^2]}\label{21}.
\end{equation}
It is well behaved both in UV and IR regions. The propagators for
$\lambda$ and $\xi$ are proportional to the inverse of
\begin{equation} 
\frac{N}{2}\int \frac{d^D k}{(2\pi)^D}
\cos^2(k\wedge p)\frac{1}{[(k+p)^2-M^2] [k^2-M^2]},
\label{22}
\end{equation}
and
\begin{equation}
{N} \frac{(\not \! p + 2 M)}2\int \frac{d^D k}{(2\pi)^D}
{\cos^2(k\wedge p)} \frac{1}{[(k+p)^2-M^2][k^2-M^2]},\label{23}
\end{equation}
respectively. They are also well behaved in UV and IR regions. 

The degree of superficial divergence
for a generic 1PI graph $\gamma$ is
\begin{equation}
d(\gamma)= D - \frac{(D-1)}2N_\psi- \frac{(D-2)}2 N_\varphi-\frac{D}2
N_F- N_\sigma- \frac32 N_\xi- 2 N_\lambda,\label{24}
\end{equation}
where $N_{\cal O}$ is the number of external lines associated to the 
field ${\cal O}$. Potentially dangerous diagrams are those contributing
to the self--energies of the $\varphi$ and $\psi$ fields since, in principle,
they are quadratic and linearly divergent, respectively.
For the self-energies of $\varphi$ and $\psi$  we find that they
diverge logarithmically and they can be removed by a wave function
renormalization of the respective field. The same happens for the
auxiliary field $F$. The renormalization factors for them are the same
so supersymmetry is preserved in the noncommutative theory. 
This analysis can be extended to the n-point functions. In $2+1$
dimensions we find nothing new showing the renormalizability of the
model at leading order of $1/N$. However, in $1+1$ dimensions there
some peculiarities. Since the scalar field is dimensionless in $1+1$
dimensions any graph involving an arbitrary number of external
$\varphi$ lines is quadratically divergent. In the four-point function
there is a partial cancellation of divergences but a logarithmic
divergence still survives. The counterterm needed to remove it can not
be written in terms of $\int d^2 x \,\,\varphi_i \star \varphi_i \star
\varphi_j\star\varphi_j$ and $\int d^2 x \,\,
\varphi_i\star\varphi_j\star\varphi_i\star\varphi_j$. A possible way 
to remove this divergence is by generalizing the definition of 1PI
diagram along the lines suggested in \cite{Arefeva1} for the
commutative nonlinear sigma model. However the cosine factors do not
allow us to use this mechanism which casts doubt about the
renormalizability of the noncommutative supersymmetric $O(N)$
nonlinear sigma model in $1+1$ dimensions. 

\section{Conclusions}

We have shown that it is possible to build consistent quantum field
theories in noncommutative space. It seems that supersymmetry is
an essential ingredient for renormalizability. The models studied here
do not involve gauge fields and this considerably simplifies the
situation. All vertices are deformed in the same way by the Moyal
product and this was essential to analyze the amplitudes. With gauge
fields the situation is much more complicated because the vertices are
deformed in different ways. However,
supersymmetric gauge theories may still have a better behavior. 

\section{Acknowledgments}
This work was done in collaboration with H. O. Girotti, M. Gomes and
A. J. da Silva. It was partially supported by Funda\c c\~ao de Amparo \`a
Pesquisa do Estado de S\~ao Paulo (FAPESP), Conselho Nacional de
Desenvolvimento Cient\'\i fico e Tecnol\'ogico (CNPq), and PRONEX
under contract CNPq 66.2002/1998-99.

\end{document}